\documentclass[aps,prl,twocolumn,showpacs,amsmath]{revtex4}
 \usepackage{hyperref}
 \usepackage{graphics} 
 \usepackage{epsfig}
\usepackage[usenames]{color}
\definecolor{blue}{rgb}{0,0.08,0.45}
   \begin{document} 
\title{
Shot noise of a mesoscopic two-particle collider 
} 
\author{
S. Ol'khovskaya$^{1}$, 
J. Splettstoesser$^{2}$, 
M. Moskalets$^{1,2}$, 
and M. B\"uttiker$^{2}$
}
\affiliation{
$^1$Department of Metal and Semic. Physics, NTU "Kharkiv Polytechnic Institute", 61002 Kharkiv, Ukraine\\
$^2$D\'epartement de Physique Th\'eorique, Universit\'e de Gen\`eve, CH-1211 Gen\`eve 4, Switzerland
}
\date\today
\begin{abstract}
We investigate the shot noise generated by particle emission from a mesoscopic capacitor into an edge state reflected and transmitted at a quantum point contact (QPC). For a capacitor subject to a periodic voltage the resulting shot noise is proportional to the number of particles (both electrons and holes) emitted during a period.  It is proportional to the product of transmission and reflection probability of the QPC independent of the applied voltage but proportional to the driving frequency.  
If two driven capacitors are coupled to a QPC at different sides then the resulting shot noise is maximally the sum of noises produced by each of the capacitors. 
However the noise is suppressed depending on the coincidence of the emission of two particles of the same kind. 
\end{abstract}
\maketitle

A prominent feature of mesoscopic systems is their ability to show quantization phenomena. 
These are, for instance, the Hall resistance in the integer \cite{intqhe} and fractional \cite{fracqhe} quantum Hall effect, and the conductance quantization of a ballistic quantum point contact \cite{ballistic}.
These quantization phenomena are governed by the number of elementary conduction channels. 
In contrast, the quantization of the charge relaxation resistance, $R_{q}$, of a quantum capacitor, predicted theoretically \cite{BTP93} and confirmed in experiment \cite{Gabelli06}, relies on the property of a single, possibly interacting, scattering channel \cite{NLB07,indWWG07}.

Of high interest are dynamic current quantization phenomena.
This quantization is governed by the number of particles participating in the transport during some fixed time interval (e.g., the driving period of a pump \cite{pump}). 
A quantized dc current was experimentally observed in a Coulomb blockade turnstile \cite{CBpumps}, in a one-dimensional channel under the action of surface acoustic waves \cite{Talyanskii97}, and recently in a 1D channel subject to either two local potentials oscillating out of phase \cite{2par_pump} or a single oscillating potential \cite{1par_pump}. 
Importantly a quantized ac current generated by a quantum capacitor subject to large amplitude excitation was observed \cite{Feve07} and discussed \cite{MSB08,KSL08}. 

These phenomena deal with the measurement of single particle observables, like the current. 
We show in this Letter that the noise, essentially a two-particle phenomenon, can exhibit a quantization behavior as well.

\begin{figure}[t]
  \centering%
  \includegraphics[width=8cm]{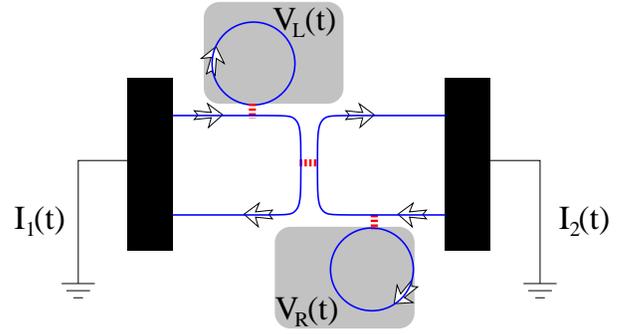}
  \caption{ (color online) 
Two quantum capacitors (circular edge states) are coupled to linear edge states which in turn are coupled, via quantum point contacts shown as short red dashed lines. Edge states are shown as blue lines with arrows indicating the direction of motion. The potentials $V_{L/R}(t)$ induced by back-gates (hatched areas) acting on the capacitors generate ac currents $I_{1/2}(t)$ at leads (black rectangles). 
}
\label{fig1}
\end{figure}

We consider the system, Fig.\,\ref{fig1}, consisting of two quantum capacitors connected to different linear edge states which in turn are coupled via a central quantum point contact (QPC). 
In the regime when either one of the quantum capacitors (or both) generate a quantized ac current \cite{Feve07} induced by an oscillating back-gate potential the shot noise, as we show, is quantized. 
If the transmission $T_{\alpha}$ of a QPC connecting the capacitor $\alpha = L,\,R$ to the linear edge state is small, $T_{\alpha} \to 0$, and the amplitude of the driving potential $V_{\alpha}(t) = V_{\alpha,0} + V_{\alpha,1}\cos(\Omega t + \varphi_{\alpha})$ is large compared to the level spacing $\Delta_{\alpha}$ then for small frequency $n_{\alpha} = [2V_{\alpha,1}/\Delta_{\alpha}]$ electrons (here $[X]$ is the integer part of $X$) and $n_{\alpha}$ holes are emitted during a driving period ${\cal T} = 2\pi/\Omega$. 
We show that, if the emission of particles is not simultaneous, the zero-frequency correlator ${\cal P}_{12}$ of currents flowing into the leads 1 and 2 is
\begin{equation}
 \label{n_01}
{\cal P}_{12} = - N {\cal} {\cal P}_{0},\, 
\end{equation} 
where 
$N = 2n_{L} + 2n_{R}$ is the total number of particles (electrons and holes) emitted during a driving period,
${\cal P}_{0} = (2e^2/h) T_{C} R_{C} \hbar\Omega $, with $T_{C}, R_{C}$ being transmission and reflection probabilities of the central QPC connecting the two linear edge states, see Fig.\,\ref{fig1}.
Note that the noise produced by the source $\alpha$ alone is: ${\cal P}_{\alpha,12} = - 2n_{\alpha} {\cal P}_{0}$.

If two electrons (or two holes) emitted by different sources arrive at the central QPC at the same time then the noise will be suppressed. 
The difference $\delta {\cal P}_{12}$ between the noise ${\cal P}_{12}$ produced by the system of two sources and the sum of noises $ {\cal P}_{L,12} + {\cal P}_{R,12}$ produced by either of the sources depends on the difference of times $\Delta t$ when particles arrive at the QPC. 
Each pair of particles arriving at the central QPC with a small time delay $\Delta t$ leads to a noise suppression, $\delta {\cal P}_{12} = {\cal P}_{12} - {\cal P}_{L,12} - {\cal P}_{R,12}$, having the form of a Breit-Wigner resonance as a function of $\Delta t$, 
\begin{equation}
 \label{01}
\frac{ { \delta\cal P}_{12} }{ { 2\cal P}_{0} }  =  \frac{ 4 \Gamma_{L} \Gamma_{R} }{ \Delta t^2 +  (\Gamma_{L} + \Gamma_{R} )^2  } \,.
\end{equation} 
Here $\Gamma_{\alpha}$ is a width in time of an electron wave packet emitted by the source $\alpha$. 
In the case of identical capacitors driven by the same potential the noise will be suppressed down to zero, ${\cal P}_{12} = 0$.

Now we show how Eqs.\,(\ref{n_01}) and (\ref{01}) were obtained. 
The system, Fig.\,\ref{fig1}, includes two single electron sources (SES) consisting of a circular edge state with length $L_{\alpha}$ connected via the QPC $\alpha$ to a corresponding linear edge state. 
The nearby gate induces a uniform potential $V_{\alpha}(t)$ over a circular state. 
The left and right linear edge states are connected to leads $1$ and $2$, respectively, which have the same chemical potential $\mu$ and the temperature $T$. 
In addition the linear edge states are coupled via the central QPC with transmission probability $T_{C}$.
The model of a SES was treated in Refs.\,\onlinecite{Gabelli06,Feve07}. 
To calculate the current and noise we use the Floquet scattering matrix 
for a SES presented for a multilevel capacitor under large sinusoidal voltage in \cite{MSB08}. 
Recently the scattering
matrix was presented for a single level
dot subject to linear in time driving [15].
In the adiabatic limit, $\Omega\to 0$, which we will consider, the elements of the Floquet scattering matrix of the SES $\alpha$ can be expressed in terms of the Fourier coefficients of the frozen scattering amplitude
\begin{equation}
 \label{03}
S_{\alpha}(t,E) = e^{ i\theta_{\alpha} } \frac{ \sqrt{1 - T_{\alpha}} - e^{i\phi_{\alpha}(t,E)} }{ 1 - \sqrt{1 - T_{\alpha}} e^{i\phi_{\alpha}(t,E)} }\,, 
\end{equation} 
where 
$\theta_{\alpha}$ is the phase of the reflection amplitude of the QPC $\alpha$;
$\phi_{\alpha}(t,E) = \phi_{\alpha}(\mu) + 2\pi\Delta_{\alpha}^{-1}[ E-\mu - eV_{\alpha}(t) ] $ is the phase accumulated by an electron with energy $E$ during one trip along the circular edge state,  
$\Delta_{\alpha} = h v_{D,\alpha}/L_{\alpha} $ with $v_{D,\alpha}$ being a drift velocity.

The frozen scattering matrix $\hat S(t,E)$ for the whole system is a $2\times 2$ unitary matrix
\begin{equation}
 \label{02}
\hat S = e^{i\theta_{C}}\left(
\begin{array}{cc}
 S_{L} e^{ikd_{L}} \sqrt{R_{C}} &
 iS_{R} e^{ikd_{R}} \sqrt{T_{C}} \\
\ \\
 iS_{L} e^{ikd_{L}} \sqrt{T_{C}} &
 S_{R} e^{ikd_{R}} \sqrt{R_{C}} 
\end{array} 
\right)\,.
\end{equation}  
Here 
$\theta_{C}$ and $R_{C}$ define the reflection amplitude, $r_{C} = \sqrt{R_{C}} e^{i\theta_{C}}$, of the central QPC;
$T_{C} = 1 - R_{C}$;
$k$ is the wave number for an electron with energy $E$; 
$d_{\alpha}$ is the distance between the source $\alpha$ and the central QPC.

We find the current $I_{j}(t)$ flowing into the contact $j = 1,\,2$ as a sum of currents produced by each of the SESs and partitioned at the central QPC: $I_{1}(t) = R_{C} I_{L}(t) + T_{C} I_{R}(t)$, and $I_{2}(t) = T_{C} I_{L}(t) + R_{C} I_{R}(t)$.
At slow driving the current can be expressed in terms of a time-dependent density of states 
$\nu_{\alpha}$ 
of a SES 
\begin{equation}
 \label{05}
\begin{array}{c}
I_{\alpha}(t) = e \int dE \left( -\frac{\partial f_{0}}{\partial E} \right) \nu_{\alpha}(t,E) \frac{d }{dt}eV_{\alpha } \,, \\
\nu_{\alpha}(t,E) = \frac{1}{\Delta_{\alpha}} \frac{T_{\alpha} }{2 - T_{\alpha} -2\sqrt{1-T_{\alpha} } \cos(\phi[t,E]) }\,,
\end{array}
\end{equation} 
where $f_{0}(E)$ is the Fermi distribution function.
We see, that the current is additive and it is not sensitive to the presence or absence of collisions of electrons emitted by different sources at the central QPC.  

In contrast the noise is sensitive to such collision processes.
We calculate the symmetrized cross-correlator of currents flowing into the leads (see, e.g., Ref.\,\onlinecite{BB00}):
$P_{12}(t,t^{\prime}) = (1/2)\left\{\langle \hat I_{1}(t) \hat I_{2}(t^{\prime}) + \hat I_{2}(t^{\prime}) \hat I_{1}(t) \rangle\right\}$,
where $\hat I_{j}(t)$ is the current operator and $\langle\cdots\rangle$ denotes averaging over the equilibrium state of the leads. 
If the driving frequency $\Omega$, the measurement frequencies $\omega$, $\omega^{\prime}$ and the temperature $T$ are all smaller than the energy scale over which the scattering matrix $\hat S$ changes significantly then the frequency representation \cite{MB07} of $P_{12}$ is 
\begin{equation}
 \label{07}
\begin{array}{l}
P_{12}(\omega,\omega^{\prime}) = \sum\limits_{l=-\infty}^{\infty} \pi {\cal P}_{12}^{(l)}(\omega) \delta(\omega + \omega^{\prime} - l\Omega)\,, \\
{\cal P}_{12}^{(l)}(\omega) = - \frac{e^2}{2\pi} T_{C} \bigg\{ \delta_{l,0} \left( \chi(\omega)  + \chi(l\Omega - \omega) \right) \\
+ R_{C} \sum\limits_{q=-\infty}^{\infty} \Big( - 2\delta_{l,0}\delta_{q,0} + (S_{L}S_{R}^{\star})_{l-q} (S_{R}S_{L}^{\star})_{q} \\
+ (S_{R}S_{L}^{\star})_{l-q} (S_{L}S_{R}^{\star})_{q}  \Big) \chi((l-q)\Omega - \omega)
\bigg\}\,,
\end{array} 
\end{equation} 
where the lower indices $q$ and $l-q$ denote the Fourier coefficients; $\chi(\omega) = \omega\coth(\hbar \omega/2k_{B}T)$.
The scattering amplitudes $S_\alpha$ are calculated at the Fermi energy, $S_{\alpha} \equiv S_{\alpha}(t,\mu)$. 

Already here we can make a general statement.
For a symmetric setup, $S_{L}(t) = S_{R}(t)$, the noise produced by the two SESs completely vanishes, 
since $S_{\alpha}S_{\alpha}^{\star} = 1$.
Only the noise of the central QPC remains. 
As there is no dc bias, the noise of the central QPC is the thermal and quantum equilibrium noise only. 

However for a non-symmetric setup, $S_{L}(t) \ne S_{R}(t)$, there is a nonequilibrium noise. 
We are interested in the noise due to the SESs only. 
Therefore, from now on we will consider the zero-frequency noise at zero temperature, where the equilibrium noise vanishes. 
From Eq.\,(\ref{07}) we get  (${\cal P}_{12}^{(0)}(0) \equiv {\cal P}_{12} $),
\begin{equation}
 \label{08}
{\cal P}_{12} = - {\cal P}_{0} \sum\limits_{q=1}^{\infty} q \left\{ \left| (S_{L}^{\star} S_{R} )_{q} \right|^2 + \left| (S_{L}^{\star} S_{R} )_{-q} \right|^2 \right\}\,.  
\end{equation}  
We distinguish a weak and a strong amplitude regime depending on whether the driving amplitude $V_{\alpha,1}$ is small or large compared to a  corresponding level spacing $\Delta_{\alpha}$.

In the weak amplitude regime, $V_{\alpha,1} \ll \Delta_{\alpha}$, the dominant contribution to noise is a bilinear function of $V_{\alpha,1}$ and it depends essentially on the density of states $\nu_{\alpha}$ (calculated at $V_{\alpha,0}$) of the SESs.
We find, ${\cal P}_{12} = {\cal P}_{L,12} + {\cal P}_{R,12} + \delta{\cal P}_{12}$, with
\begin{equation}
 \label{17}
\begin{array}{l}
{\cal P}_{\alpha,12} = -\frac{1}{2}{\cal P}_{0}(2\pi \nu_{\alpha} e V_{\alpha,1})^2\,, \\
\ \\
\delta {\cal P}_{12} = 4\pi^2e^2{\cal P}_{0}\nu_{L} \nu_{R} V_{L,1}  V_{R,1} \cos( \Delta\varphi )\,.
\end{array}
\end{equation} 
Here $\Delta\varphi = \varphi_{L} - \varphi_{R}$ is the phase lag between the two sources.
In this regime, in general, the sources produce different noises, ${\cal P}_{L,12} \ne {\cal P}_{R,12}$. 
The correlation contribution to noise, $\delta{\cal P}_{12}$, being proportional to $\cos(\Delta\varphi)$ can either enhance or suppress the total noise produced by two sources. 
In the weak amplitude regime the noise is a consequence of electron-hole pairs generated by the periodic potential acting on the quantum capacitor.  
Physically it is similar to the generation of shot noise by an ac voltage applied across a QPC investigated experimentally in Ref.\,\cite{RRGEJ03}. 
In both cases there is a correlation contribution to the noise \cite{RPB05}.

The noise properties are completely different in the strong amplitude regime, $V_{\alpha,1} \gtrsim \Delta_{\alpha}$, when the SES can emit an electron or absorb an electron (i.e., emit a hole).
This happens periodically each time when one of the quantum levels of an SES crosses the Fermi level of a lead. 
The width of a resulting current pulse depends on the transparency $T_{\alpha}$ of the QPC connecting circular and linear edge states. 
In the limit of a small transparency, $T_{\alpha}\to 0$, the current pulse is narrow on the scale of a period ${\cal T} = 2\pi/\Omega$. 
In this case an individual additional electron (hole) propagates through the system. 
Since there are two sources one can get situations with either one or two particles propagating through the system at the same short time period.
If two particles propagate through the system the noise will be suppressed, due to the Pauli exclusion principle. 
If electrons (holes) meet each other at the same place (at the central QPC) they become anticorrelated.
This anticorrelation completely suppresses the partition noise which would arise if they would pass the central QPC at different times.
This opens the possibility to use such a double capacitor system as an on-demand source of entangled pairs of particles \cite{pumpent}.

To proceed analytically we assume that the amplitude of an oscillating potential is chosen in such a way that during the period only one level of the SES crosses the Fermi level. The time of crossing $t_{0,\alpha}$ is defined by $\phi_{\alpha}(t_{0,\alpha}) = 0 \mod 2\pi$.
Introducing the deviation of the phase from its resonance value, $\delta\phi_{\alpha}(t) = \phi_{\alpha}(t) - \phi_{\alpha}(t_{0,\alpha}) $, we obtain the scattering amplitudes, Eq.\,(\ref{03}), in the limit $T_{\alpha}\to 0$ as follows
\begin{equation}
 \label{09}
S_{\alpha}(t) = - e^{i\theta_{\alpha}} \frac{T_{\alpha} + 2i\delta\phi_{\alpha}(t) }{T_{\alpha} - 2i\delta\phi_{\alpha}(t)} + {\cal O} (T_{\alpha}^{2}) \,.
\end{equation}  
We keep only terms in leading order in $T_{\alpha}$. 

There are two time moments when resonance conditions occur. 
One time is when the level sinks below the Fermi level and the second one is when the level rises above the Fermi level.
We will denote these times as $t_{0,\alpha}^{(-)}$ and $t_{0,\alpha}^{(+)}$, respectively.
At time $t_{0,\alpha}^{(-)}$ one electron is emitted by the source $\alpha$, while at time 
$t_{0,\alpha}^{(+)}$ one electron enters the dot (a hole is emitted). 

We suppose that the constant part of the potential 
$ -\Delta_{\alpha}/2 < eV_{\alpha,0} < \Delta_{\alpha}/2 $ accounts for a detuning of an electron level in the SES from the Fermi level. 
Then for $|eV_{0,\alpha}| < e V_{1,\alpha} < \Delta_{\alpha} - |eV_{0,\alpha}|$ we get the resonance times, 
$\Omega t_{\alpha}^{(\mp)} = \mp \arccos\left( - V_{0,\alpha} /V_{1,\alpha} \right) - \varphi_{\alpha}$.
The deviation $\delta t_{\alpha}^{\mp} = t - t_{\alpha}^{\mp}$ can be related to a deviation from the resonance phase,
$ \delta\phi_{\alpha} = \mp M_{\alpha} \Omega \delta t_{\alpha}^{\mp}$, 
where $\mp M_{\alpha}= d\phi_{\alpha}/dt|_{t = t_{\alpha}^{\mp}}/\Omega = \mp 2\pi |e| \Delta_{\alpha}^{-1} \sqrt{V_{1,\alpha}^{2} - V_{0,\alpha}^{2} }\,$.
With these definitions we can rewrite Eq.\,(\ref{09}) assuming that the overlap between the resonances is small, $t_{\alpha}^{(+)} - t_{\alpha}^{(-)} \gg T_{\alpha}/\Omega$,
\begin{equation}
 \label{13}
S_{\alpha}(t) \approx e^{i\theta_{\alpha}} \left\{
\begin{array}{cc}
\frac{ t - t_{\alpha}^{(+)}  + i \Gamma_{\alpha}   }{t - t_{\alpha}^{(+)}  - i \Gamma_{\alpha}   }\,,  &
0 < t + \frac{\varphi_{\alpha} }{\Omega } <  \frac{{\cal T} }{2}\,, \\
\ \\
\frac{ t - t_{\alpha}^{(-)} - i \Gamma_{\alpha} }{t - t_{\alpha}^{(-)}  + i \Gamma_{\alpha} }\,, &
\frac{{\cal T} }{2} < t + \frac{\varphi_{\alpha} }{\Omega } < {\cal T}\,.
\end{array}
\right.
\end{equation} 
where $\Gamma_{\alpha} = T_{\alpha}/(2\Omega M_{\alpha})$. 
This scattering amplitude leads to the time-dependent current $I_{\alpha}(t)$, Eq.\,(\ref{05}), generated by the source $\alpha$ for $0 < t < {\cal T}$,
\begin{equation}
 \label{14}
\!\!\!\!I_{\alpha}(t) =\! \frac{e}{\pi}\! \left\{\! \frac{\Gamma_{\alpha} }{ \left( t - t_{\alpha}^{(-)} \right)^2 + \Gamma_{\alpha}^{2} } - \frac{\Gamma_{\alpha} }{ \left( t - t_{\alpha}^{(+)} \right)^2 + \Gamma_{\alpha}^{2} } \!\right\}.
\end{equation} 
This current consists of two pulses of the same width $\Gamma_{\alpha}$ corresponding to an emission of an electron and a hole.

Now we can calculate the zero-frequency noise power. 
First we calculate the noise ${\cal P}_{\alpha,12}$ produced by only one of the sources if the second source is stationary. 
Substituting Eq.\,(\ref{13}) into Eq.\,(\ref{08}) we find the noise
\begin{equation}
 \label{15}
{\cal P}_{L,12} = {\cal P}_{R,12} = - 2 {\cal P}_{0}\,.
\end{equation}
which is independent of the parameters of the source. 
This noise can be understood as the shot noise produced by the central QPC under the action of one electron and one hole emitted by the source during a period ${\cal T} = 2\pi/\Omega$.  
Since electron and hole are emitted at different times they are uncorrelated and contribute to the noise independently.
Since the electron-hole symmetry is not violated in our system they contribute to the noise equally, leading to a factor 2 in Eq.\,(\ref{15}). 
The noise, ${\cal P}_{0}$, produced by the single electron (hole) pulse in our case coincides with the shot noise of the central QPC (having only one conducting channel)  which is subject to a time-independent voltage $|eV| = \hbar\Omega$, ${\cal P}_{12}^{(dc)} = - (2e^2/h) T_{C} R_{C} |eV|$ \cite{BB00}.  

\begin{figure}[t]
  \centering%
   \includegraphics[width=8cm]{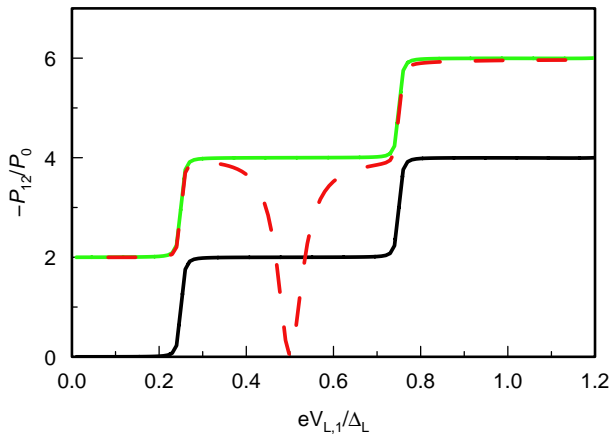}
  \caption{ (color online) The noise ${\cal P}_{12}$, Eq.\,(\ref{08}),  as a function of the amplitude $V_{L,1}$ of the potential acting upon the left capacitor. Black (lower) solid line: the right capacitor is stationary, $V_{R,1} = 0$. Green (upper) solid line: the right capacitor is driven by the out of phase potential, $\varphi_{R} = \pi$, with amplitude $V_{R,1} = 0.5\Delta_R$. Red dashed line: the right capacitor is driven by the in phase potential, $\varphi_{R} = 0$, with amplitude $V_{R,1} = 0.5\Delta_{R}$. The parameters are: $V_{L,0} = V_{R,0} =0.25\Delta_{R}$ ($\Delta_{L} = \Delta_{R}$), $\varphi_{L} = 0$. The noise is given in units of ${\cal P}_{0} = (2e^2/h) T_{C} R_{C} \hbar\Omega$. 
}
\label{fig2}
\end{figure}

If the amplitude of driving is larger, for instance if $n$ electrons and $n$ holes are emitted during a period, then the noise is $n$ times larger, ${\cal P}_{\alpha,12} = - 2n{\cal P}_{0}$, as shown in Fig.\,\ref{fig2}, black (lower) solid line. 
Remarkably the noise produced by the SES is quantized.
The increment ${\cal P}_{0}$, Eq.\,(\ref{15}), depends on the frequency $\Omega$ of the oscillating voltage and on the transparency $T_{C}$ of the central QPC. 
Therefore the quantization is not universal. 

The noise of the system of two sources depends crucially on whether particles are emitted at the same time
or not.
Substituting Eq.\,(\ref{13}) into Eq.\,(\ref{08}) we find 
\begin{equation}
 \label{16}
\begin{array}{c}
{\cal P}_{12} = -2 {\cal P}_{0} \bigg\{ 2 - \frac{4\Gamma_{L}\Gamma_{R} }{ \left( t_{L}^{(-)} - t_{R}^{(-)} \right)^2 + (\Gamma_{L} + \Gamma_{R})^2 } \\
- \frac{4\Gamma_{L}\Gamma_{R} }{ \left( t_{L}^{(+)} - t_{R}^{(+)} \right)^2 + (\Gamma_{L} + \Gamma_{R})^2 } + {\cal O}(\Gamma_{\alpha}^{2} ) \bigg\} \,.  
\end{array} 
\end{equation}  
Particles of the same kind (either electrons or holes), which are emitted by different sources, can arrive at the central QPC with a time difference
$\Delta t^{(\mp)} = t_{L}^{(\mp)} - t_{R}^{(\mp)}$. When this time difference  
is larger than the sum of the widths of the corresponding current pulses, $\Delta t^{(\mp)} \gg \Gamma_{L} + \Gamma_{R}$, then the two sources contribute to the noise independently, Fig.\,\ref{fig2}, green (upper) solid line. 
In this case Eq.\,(\ref{16}) leads to Eq.\,(\ref{n_01}).

In contrast if there is some overlap in time between particle wave packets, $\Delta t^{(-)} \sim \Gamma_{L} + \Gamma_{R}$ or $\Delta t^{(+)} \sim \Gamma_{L} + \Gamma_{R}$, then the correlation contribution $\delta{\cal P}_{12} $ to the noise arises. 
From Eq.\,(\ref{16}) we get Eq.\,(\ref{01}) with $\Delta t$ being either $\Delta t^{(-)}$ or $\Delta t^{(+)}$.
If $\Delta t^{(\mp)} = 0$ then the noise is maximally suppressed. 
For a fully symmetrical case, $\Gamma_{L} = \Gamma_{R}$ and $V_{L}(t) = V_{R}(t)$, the noise is suppressed down to zero, while the amplitudes of current pulses are rather enhanced.
In Fig.\,\ref{fig2} the red dashed line shows the noise generated by two equal sources as a function of the amplitude $V_{L,1}$. 
If $V_{L,1}\ne V_{R,1}$ then the times when particles are emitted by different sources are different. In this case both sources contribute to noise independently.
However if $V_{L,1}$ approaches $V_{R,1}=0.5\Delta_{R}$ then the time difference $\Delta t^{(\mp)}\to 0$ which results in suppression of the shot noise. 

One should note that an electron-hole collision at the central QPC does not affect the noise. 
Because the emitted electron has an energy above the Fermi level and the emitted hole has an energy below the Fermi level they are not subject to the Pauli exclusion principle. 
Hence they do contribute to noise independently.

In conclusion, 
we predict two phenomena. 
First, the quantum mesoscopic capacitor subject to a large amplitude voltage and connected to a QPC produces shot noise which is quantized. 
Second, the noise of two capacitors coupled in parallel to a QPC is suppressed when they emit electrons simultaneously. 
These phenomena are the basis for the design of a two-particle emitter with a controllable degree of correlations.

We thank D.C. Glattli for suggesting this problem. 
M.~B. and J.~S. acknowledge the support of the Swiss NSF and the European STREP project SUBTLE.

\end{document}